\documentclass[aps,prl,reprint,amsmath,amssymb,floatfix,citeautoscript]{revtex4-2}

\usepackage[T1]{fontenc}
\usepackage[latin9]{inputenc}
\usepackage{babel}
\usepackage{graphicx}
\usepackage{txfonts}
\usepackage{bm}
\usepackage{color}
\definecolor{myblue}{rgb}{0,0,1}
\usepackage[breaklinks=true,colorlinks=true,linkcolor=myblue,urlcolor=myblue,citecolor=myblue]{hyperref}

\newcommand{\vc}[1]{{\bm{#1}}}

\newcommand{\diff}{\mathop{}\!d}

\begin{document}


\title{Treating geometric phase effects in nonadiabatic dynamics}

\author{Alex Krotz}
\author{Roel Tempelaar}
\email{roel.tempelaar@northwestern.edu}

\affiliation{Department of Chemistry, Northwestern University, 2145 Sheridan Road, Evanston, Illinois 60208, USA}

\begin{abstract}
 We present an approach for eliminating the gauge freedom for derivative couplings in nonadiabatic dynamics in the presence of geometric phase effects. This approach relies on a bottom-up construction of a parametric quantum Hamiltonian in terms of functions of a dynamical variable, which can be associated with real and imaginary-valued contributions to the Hamiltonian in a given diabatic basis. By minimizing the deviation of the imaginary functions from a constant we identify a set of diabatic bases that recover the real-valued gauge commonly used for topologically-trivial systems. This minimization, however, also confines the gauge freedom in the topologically-nontrivial case, opening a path towards finding gauge-invariant derivative couplings under geometric phase effects. Encouraging results are presented for fewest-switches surface hopping calculations of a nuclear wavepacket traversing a single avoided crossing, for which fully gauge-invariant derivative couplings are found.
\end{abstract}

\maketitle

\emph{Introduction.}---It is well known that the parametric dependence of a quantum Hamiltonian on a dynamical variable may give rise to geometric phase effects that influence the dynamics \cite{berry1984QuantalPhaseFactors, baer2006BornOppenheimerConicalIntersections, min2014MolecularBerryPhase}. This is particularly relevant to mixed quantum-classical dynamics which involves a quantum Hamiltonian that depends parametrically on classical coordinates \cite{krishna2007PathIntegralFormulation, subotnik2019DemonstrationConsistencyQuantum}.

In topologically-trivial systems the nonadiabatic rotation of the quantum eigenbasis induced by the classical coordinates is captured by the derivative couplings between eigenstates which can be made real-valued by appropriately adjusting the gauge of the eigenbasis \cite{mead1992GeometricPhaseMolecular}. In topologically-nontrivial systems, however, the derivative coupling also captures a ``geometric'' rotation that is orthogonal to the otherwise real-valued nonadiabatic rotation. 

In addition to giving rise to a path-dependent phase-factor on individual eigenstates in the adiabatic limit \cite{berry1984QuantalPhaseFactors}, the geometric rotation introduces a complex contribution to the derivative couplings between eigenstates \footnote{We make a point of only referring to topologically-nontrivial systems as having geometric phase effects despite the fact that nonadiabatic dynamics in topologically-trivial systems can also be considered as a type of geometric rotation \cite{anandan1988NonadiabaticNonabelianGeometric}.}. As a result, the derivative couplings can no longer be made real-valued by a gauge transformation. This complicates the description of topologically-nontrivial systems by trajectory surface hopping techniques \cite{tully1971TrajectorySurfaceHopping,tully1991NonadiabaticMolecularDynamics,barbatti2011NonadiabaticDynamicsTrajectory,yu2014TrajectorybasedNonadiabaticMolecular}, such as the widely-used fewest-switches surface hopping (FSSH) \cite{tully1990MolecularDynamicsElectronic, hammes-schiffer1994ProtonTransferSolution}, where real-valued vectors associated with derivative couplings are required to determine the direction in which to rescale classical momenta upon nonadiabatic transitions (hops).

The recent years have seen a growing interest in the application of trajectory surface hopping techniques to problems featuring nontrivial topologies \cite{bian2021ModelingNonadiabaticDynamics,wu2022PhaseSpaceSemiclassicalApproach,xie2017AccuracyTrajectorySurfacehopping,bian2022IncorporatingBerryForce,mai2015GeneralMethodDescribe}, including pronounced spin-dependent behavior \cite{wu2020ChemicalReactionRates, wu2021ElectronicSpinSeparation, gohler2011SpinSelectivityElectron,naaman2012ChiralInducedSpinSelectivity,naaman2015SpintronicsChiralitySpin,naaman2019ChiralMoleculesElectron}, conical intersections \cite{xie2017NonadiabaticTunnelingConical,xie2019SignInsidiousEffects,yuan2020ObservationGeometricPhase}, intersystem crossings \cite{fedorov2016InitioMultipleSpawning,hauser1991IntersystemCrossingDynamics,richter2014UltrafastIntersystemCrossing}, and Dirac cones \cite{nie2014UltrafastCarrierThermalization,shi2020IodineSulfurVacancy,datta2019NontrivialQuantumOscillation}, prompting investigations into the influence of geometric phase effects on this class of methods, and on the momentum rescaling in particular \cite{bian2022ModelingSpinDependentNonadiabatic,miao2019ExtensionFewestSwitches,wu2021SemiclassicalDescriptionNuclear}.

Traditionally, rescaling directions have been associated with the instantaneous limit of the Pechukas force arising from a nonadiabatic transition \cite{pechukas1969TimeDependentSemiclassicalScattering, pechukas1969TimeDependentSemiclassicalScatteringa,miao2019ExtensionFewestSwitches}. In this limit, the Pechukas force becomes an impulsive one in the direction of the real part of the derivative coupling and so is real-valued by construction \cite{tully1991NonadiabaticMolecularDynamics}. Nevertheless, the direction of the real part of the derivative coupling is generally not gauge invariant. Indeed, even in the topologically-trivial case, a gauge exists where the derivative coupling becomes fully imaginary as a result of which the direction of its real part becomes undefined \footnote{An example is the case where the real-valued eigenvectors $\vert\alpha\rangle$ and $\vert\beta\rangle$ are transformed as $\vert\alpha'\rangle=\vert\alpha\rangle$ and $\vert\beta'\rangle=e^{i\frac{\pi}{2}}\vert\beta\rangle$, to yield the fully imaginary-valued derivative coupling $\langle\alpha'\vert\nabla\beta'\rangle$}.

If anything, this suggests that previous works have adopted an implicit gauge fixing which at the very least disfavors this ``singular gauge.'' An obvious choice of gauge for the topologically-trivial case is then one in which the derivative couplings become fully real-valued, consistent with the traditional application of trajectory surface hopping techniques to systems with real-valued eigenvectors. In the topologically-nontrivial case, however, there is no gauge in which the derivative couplings become fully real-valued and a change in gauge rotates their real part, changing the momentum rescaling direction. The sudden appearance of a gauge ambiguity when geometric phase effects are introduced motivates the formulation of a more general treatment that applies to both the topologically trivial and nontrivial cases, necessarily recovering real-valued derivative couplings in the former and restricting the gauge freedom in the latter.

In this Letter we pursue this goal by presenting a bottom-up construction of a parametric quantum Hamiltonian in terms of functions of a dynamical variable. These functions can be associated with real and imaginary-valued contributions to the Hamiltonian in a given diabatic basis. We then recognize that in certain ``preferred'' bases the functions associated with the imaginary-valued contributions can be taken as constants in the topologically-trivial case, consequently giving rise to strictly real-valued derivative couplings. This allows us to employ the \textit{deviation} of these functions from constants as a metric which can be minimized to yield real-valued derivative couplings thereby resolving the singular gauge ambiguity.

By employing the aforementioned metric we establish a continuity between the topologically-trivial and nontrivial cases enabling us to radically restrict the gauge freedom for complex-valued derivative couplings arising in the topologically-nontrivial case, thereby reducing the ambiguity in the direction of their real parts. Moreover, for a two-dimensional avoided crossing problem we show empirically that this approach yields fully gauge-invariant momentum rescaling directions. In addition to shedding light on the fundamental properties of derivative couplings, our theory guides the gauge-invariant implementation of trajectory surface hopping techniques in the presence of geometric phase effects and provides a path toward incorporating such effects in future mixed quantum-classical dynamics methods.

\emph{Theory.}---We begin by considering an arbitrary quantum Hamiltonian that depends parametrically on a classical coordinate $\vc{q}$. This Hamiltonian can be represented by a set of real-valued scalar functions of $\vc{q}$, an obvious choice for which are the real and imaginary parts of the Hamiltonian matrix elements. Assuming the Hamiltonian to be $N$-dimensional and traceless, we need at most $N^{2}-1$ functions to define it and can discard at least $N'=\frac{1}{2}(N^{2}-N)$ functions if the Hamiltonian is constrained to be topologically-trivial with real-valued matrix elements. 

Naturally, these functions depend on the diabatic basis used to express the Hamiltonian. Indeed, a topologically-trivial Hamiltonian that has real-valued matrix elements in one diabatic basis may have complex-valued matrix elements in another, potentially giving rise to complex-valued derivative couplings despite remaining topologically-trivial \footnote{Here we assume an implicit convention whereby real-valued Hamiltonians always give rise to real-valued eigenvectors but complex-valued Hamiltonians under the same convention give rise to complex-valued eigenvectors. This is akin to a hypothetical analytic diagonalization.}. This implies that the functional representation still captures some of the gauge freedom of the derivative couplings, including the singular gauge. However, as shown below, by means of the functions we can avoid the singular gauge by imposing that topologically-trivial Hamiltonians necessarily gives rise to real-valued derivative couplings. 

To this end we must assess the eigenvector matrix in terms of the functions. This can be achieved by analytic diagonalization of the Hamiltonian, but doing so becomes intractable in higher dimensions. Instead, we utilize the functions to directly construct a unitary matrix of column eigenvectors and a real-valued, traceless, and diagonal matrix of eigenvalues, both of which depend parametrically on $\vc{q}$ (after which the Hamiltonian can be obtained through the eigendecomposition). 

A natural way to construct this representation is by using the generators of the group SU($N$) which forms a basis of $N$-dimensional, traceless, antihermitian matrices. While the particular form of each generator is arbitrary, we broadly classify them into three subsets: $\{T_n^{E}\}$, the subset of $N-1$ imaginary-valued diagonal generators, $\{T_n^{R}\}$, the subset of $N'$ real-valued off-diagonal generators, and $\{T_n^{I}\}$, the subset of $N'$ imaginary-valued off-diagonal generators. To formulate a generic traceless eigenvalue matrix we associate each imaginary-valued diagonal generator in $T^{E}$ with an ``energetic'' function of $\vc{q}$, $\lambda_{n}$, which linearly combines the generators to produce
 \begin{equation}
 E(\vc{\lambda}) = i \sum_{n}\lambda_{n} T^{E}_{n}.\label{eq:E}
 \end{equation}
The space of all possible traceless eigenvalue matrices is spanned through Eq.~\ref{eq:E} by variations in the functions $\lambda_{n}$. If a given Hamiltonian $H_{0}$ has the eigendecomposition $H_{0}(\vc{q})=V_{0}(\vc{q})E_{0}(\vc{q})V_{0}^{\dagger}(\vc{q})$ where $E_{0}$ and $V_{0}$ are its eigenvalue and eigenvector matrices, respectively, then $\vc{\lambda}=(\lambda_{1},\lambda_{2},...,\lambda_{N-1})$ are its energetic functions if $E(\vc{\lambda}(\vc{q}))=E_{0}(\vc{q})$ is satisfied.

Next, we recognize that a given eigenvector matrix $V_{0}$ is related to a subset of SU($N$), the elements of which are related to one another by a gauge transformation. Accordingly, we can formulate a generic eigenvector matrix up to a gauge transformation as
 \begin{equation}
 V(\vc{I},\vc{R})=\prod_{n}e^{I_{n}T^{I}_{n}}\prod_{n}e^{R_{n}T^{R}_{n}}\equiv U(\vc{I})\tilde{V}(\vc{R})\label{eq:V_3}
 \end{equation}
where $\vc{I}=(I_{1},I_{2},...,I_{N'})$ are functions associated with the imaginary-valued offdiagonal generators and $\vc{R}=(R_{1},R_{2},...,R_{N'})$ are functions associated with the real-valued offdiagonal generators. We take the ordering and choice of generators to be fixed as a convention and expand on this choice of construction in the Supplemental Material (SM).

A generic traceless Hamiltonian matrix then follows from the eigendecomposition as
 \begin{equation}
 H(\vc{\lambda},\vc{I},\vc{R}) = V(\vc{I},\vc{R})E(\vc{\lambda})V^{\dagger}(\vc{I},\vc{R}),\label{eq:H}
 \end{equation} 
which, as expected, involves $N^{2}-1$ independent functions. If we discard $U$ from the construction of $V$ (Eq.~\ref{eq:V_3}), by setting $\vc{I}=0$, we can write a generic real-valued Hamiltonian as $\tilde{H}(\vc{\lambda},\vc{R}) = \tilde{V}(\vc{R})E(\vc{\lambda})\tilde{V}^{\dagger}(\vc{R})$, which requires $N'$ fewer functions. To represent a given traceless Hamiltonian $H_{0}$ using Eq.~\ref{eq:H}, one simply solves the equation $H(\vc{\lambda}(\vc{q}),\vc{I}(\vc{q}),\vc{R}(\vc{q}))=H_{0}(\vc{q})$ yielding a functional representation of $H_{0}$. 

Once a Hamiltonian has been represented in terms of functions they can be used to evaluate the derivative couplings, which follow from Eq.~\ref{eq:V_3} as
 \begin{eqnarray}
 V^{\dagger}(\vc{I},\vc{R})\nabla V(\vc{I},\vc{R}) & = & \tilde{V}^{\dagger}({\vc{R}})U^{\dagger}(\vc{I})\nabla U(\vc{I})\tilde{V}(\vc{R})\nonumber\\
 & & +\tilde{V}^{\dagger}(\vc{R})\nabla\tilde{V}(\vc{R}),\label{eq:VdV_2}
 \end{eqnarray}
where $\nabla$ denotes differentiation with respect to $\vc{q}$ and acts only on its adjacent term. 

In Eq.~\ref{eq:VdV_2}, the term $\tilde{V}^{\dagger}({\vc{R}})U^{\dagger}(\vc{I})\nabla U(\vc{I})\tilde{V}(\vc{R})$ is generally complex-valued and vanishes when $\vc{I}=0$. It therefore does not contribute to the derivative coupling when the Hamiltonian is topologically-trivial and real-valued. More generally, it does not contribute when $\nabla\vc{I}=0$ \footnote{Because the equation $H(\vc{\lambda}(\vc{q}),\vc{I}(\vc{q},\vc{R}(\vc{q})=H_{0}(\vc{q})$ may have multiple solutions, we refer to properties of the entire set of solutions when requiring that $\nabla \vc{I}\neq 0$, such that if even one solution has $\nabla \vc{I}=0$, Eq.~\ref{eq:VdV_2} becomes real-valued.}. Indeed, for topologically-trivial Hamiltonians with $\nabla\vc{I}=0$ the nonadiabatic rotation of the eigenbasis is entirely captured by $\tilde{V}^{\dagger}(\vc{R})\nabla\tilde{V}(\vc{R})$, which is real-valued by construction. As a result, it is appropriate to use this term to determine momentum rescaling directions. 

However, as mentioned, a topologically-trivial Hamiltonian can be made complex-valued by a diabatic basis transformation, as a result of which $\nabla\vc{I}\neq0$ and $\tilde{V}^{\dagger}(\vc{R})\nabla\tilde{V}(\vc{R})$ no longer exclusively incorporates the nonadiabatic rotation. Importantly, however, in such cases, one can always find a diabatic basis for which $\nabla\vc{I}=0$. This basis is part of a set of ``preferred bases'' for which $\tilde{V}^{\dagger}(\vc{R})\nabla\tilde{V}(\vc{R})$ optimally captures the nonadiabatic rotation. Hence, in practice one may solve $H(\vc{\lambda}(\vc{q}),\vc{I}^{k}(\vc{q}),\vc{R}^{k}(\vc{q}))=D^{\dagger}_{k}H_{0}(\vc{q})D_{k}$ for arbitrary diabatic transformations, where $\vc{R}^{k}$, $\vc{I}^{k}$, and $\vc{\lambda}$ are the functions representing the Hamiltonian associated with $D_{k}$, while minimizing the deviation of $\vc{I}^{k}$ from being a constant. We note that $\vc{\lambda}$ is independent of the diabatic basis and therefore does not carry a $k$-dependence.

For simplicity, consider the case where $\vc{I}^{k}$ consists of only a single function $I^{k}$ which depends on a single coordinate $q$. The deviation of $I^{k}$ from being a constant can be assessed in terms of its Taylor series as
 \begin{equation}
 I^{k}(q+\epsilon)-I^{k}(q) = \sum_{n=1}^{\infty} \frac{1}{n!}\frac{\partial^{n}I^{k}(q)}{\partial q^{n}}\epsilon^{n},\label{eq:const_1}
 \end{equation}
which for any constant $I^{k}$ is exactly zero for any displacement $\epsilon$. By taking $I^{k}$ to be a vector in the mononomial basis in orders of $\epsilon$, the coefficients of Eq.~\ref{eq:const_1} can be considered elements of the vector describing the displacement of $I^{k}$ from the point of constancy, defined by the zeroth order element which is subtracted out in Eq.~\ref{eq:const_1}. The Euclidean distance of $I^{k}$ from this point,
 \begin{equation}
 \mathrm{dist}(I^{k}(q)) = \sqrt{\sum_{n=1}^{\infty}\left(\frac{1}{n!}\frac{\partial^{n}I^{k}(q)}{\partial q^{n}}\right)^{2}},\label{eq:dist_1}
 \end{equation}
then provides a metric for the extent to which $I^{k}$ is locally constant. 

A global metric can then be obtained by integrating Eq.~\ref{eq:dist_1} over the phase-space of $q$
 \begin{equation}
 d(I^{k}) = \Omega^{-1}\int \mathrm{dist}(I^{k}(q))\diff q,\label{eq:dist_2}
 \end{equation}
and normalizing by the area of $q$ integrated, $\Omega$. For multidimensional coordinates $\vc{q}$, Eq.~\ref{eq:dist_1} is trivially generalized to include the terms of the multidimensional Taylor series, and for multiple functions $\vc{I}^{k}=(I_{1}^{k},I_{2}^{k},...,I^{k}_{N'})$ the individual contributions to the distance can be directly summed.

Solving $H(\vc{\lambda}(\vc{q}),\vc{I}^{k}(\vc{q}),\vc{R}^{k}(\vc{q}))=D^{\dagger}_{k}H_{0}(\vc{q})D_{k}$ while minimizing $d(\vc{I}^{k})$ across diabatic bases not only recovers fully real-valued derivative couplings for the topologically-trivial case, but also provides a means for extending the concept of preferred bases to topologically-nontrivial Hamiltonians. While there is no diabatic basis in which a topologically-nontrivial Hamiltonian has $\nabla\vc{I}(\vc{q})=0$, taking the preferred bases to be those which minimize the metric $d(\vc{I}^{k})$ provides an extension which continuously reduces to the topologically-trivial case, offering a consistent framework for decomposing the derivative couplings through Eq.~\ref{eq:VdV_2}. The purely-real contribution, $\tilde{V}^{\dagger}(\vc{R})\nabla\tilde{V}(\vc{R})$, then serves to determine the momentum rescaling direction. 

Specifically for the $N=2$ case, an arbitrary traceless Hamiltonian in some given diabatic basis takes the form
 \begin{equation}
 H_{0}(\vc{q})= \rho(\vc{q})\left(\begin{array}{cc}
 -\cos \theta(\vc{q}) & \sin\theta(\vc{q})e^{i\phi(\vc{q})} \\
 \sin\theta(\vc{q})e^{-i\phi(\vc{q})} & \cos\theta(\vc{q})
 \end{array}\right),\label{eq:H_01}
 \end{equation} 
which involves only a single generator for each of the three subsets, and hence a single function for each. Evaluating $d(I^{k})$ for the solutions of $H(\lambda(\vc{q}),I^{k}(\vc{q}),R^{k}(\vc{q}))=D_{k}^{\dagger}H_{0}(\vc{q})D_{k}$ enables one to find the set of bases in which the metric is minimized. We demonstrate the construction of $H$ according to Eq.~\ref{eq:H} in the SM.

\emph{Results.}---We now proceed with an application of our approach to the example of a nuclear wavepacket traversing a single avoided crossing, for which Eq.~\ref{eq:H_01} is parameterized as
 \begin{equation}
 \rho = A, \quad
 \theta = \frac{\pi}{2}(\mathrm{erf}(B x)+1), \quad
 \phi = W y.\label{eq:params_1}
 \end{equation}
Here, $A$, $B$, and $W$ are constants, and $\vc{q}=(x,y)$. We can compute the metrics of all transformed diabatic bases by making use of the aforementioned procedure. A comprehensive survey of these metrics and the associated momentum rescaling directions is included in the SM for the values $A=20$, $B=1.0$, and either $W=0.0$, $W=2.0$, or $W=3.0$, where $W=0.0$ corresponds to the topologically-trivial case.

In one particular diabatic basis $D_{\bar{k}}$ which minimizes the metric (chosen only for notational simplicity of the associated functions) the functional representation is found to be
 \begin{equation}
 \lambda = A, \quad
 R^{\bar{k}} = -\frac{\pi}{4}\mathrm{erf}(B x), \quad
 I^{\bar{k}} = \frac{1}{2}W y.\label{eq:functions_1}
 \end{equation} 
The associated metric is
 \begin{equation}
 d(I^{\bar{k}}) = \frac{1}{2}\vert W\vert,\label{eq:min_1}
 \end{equation} 
with the rescaling direction governed by
 \begin{equation}
 \tilde{V}^{\dagger}(R^{\bar{k}})\nabla\tilde{V}(R^{\bar{k}})=-\frac{\sqrt{\pi}}{2} B e^{-B^{2}x^{2}}\hat{x} T^{R},\label{eq:nad_1}
 \end{equation}
where $\hat{x}$ denotes the unit vector in the $x$ direction and $T^{R}$ is the real-valued offdiagonal generator of SU($2$) provided in the SM. From the survey presented in the SM, it follows that Eq.~\ref{eq:min_1} is a minimum shared by a set of preferred bases that extends beyond $\bar{k}$, of which all yield the same momentum rescaling direction up to an overall sign. This implies that within trajectory surface hopping techniques momentum should be unambiguously rescaled in the $x$ direction.

\begin{figure}
\centering
\includegraphics{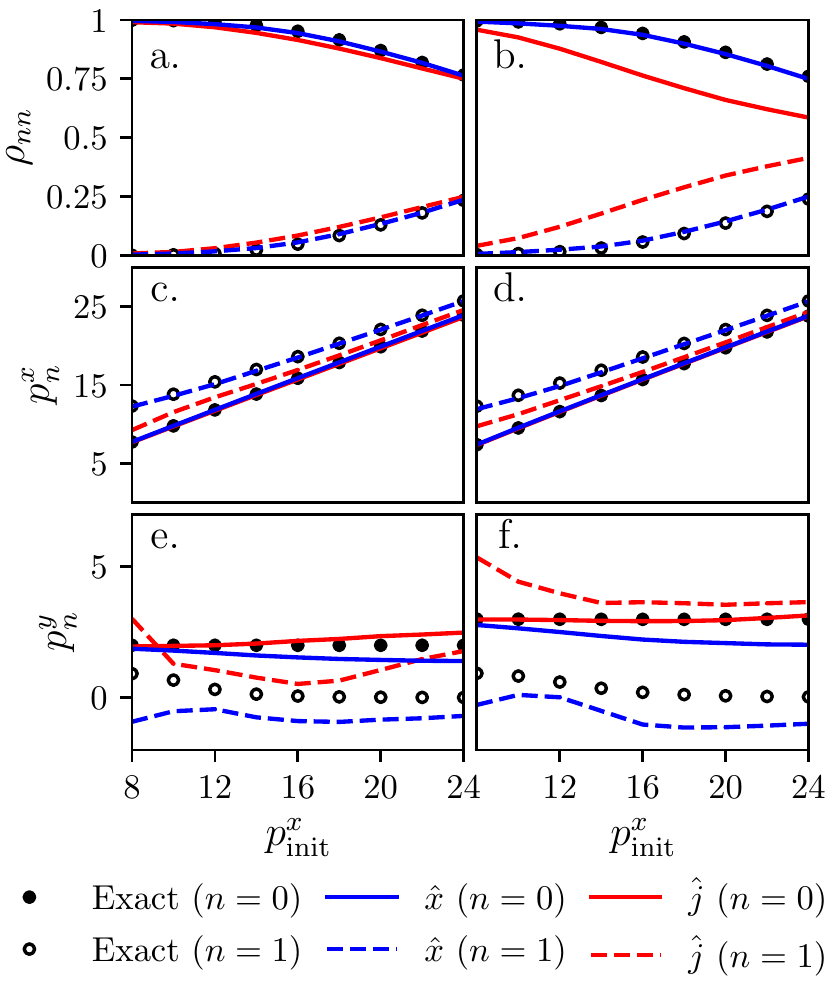}
\caption{Comparison of FSSH calculations incorporating different momentum rescaling directions against exact results, for a nuclear wavepacket traversing a single avoided crossing. Shown are the transmitted population $\rho_{nn}$ and the $x$ and $y$ components of momentum ($p^{x}_{n}$ and $p^{y}_{n}$, respectively) for the upper ($n=1$) and lower ($n=0$) diabatic surfaces with $B=1.0$, and $W=2.0$ (a,c,e) or $W=3.0$ (b,d,f). Rescaling along $\hat{x}$ is consistent with the approach outlined in this Letter, whereas rescaling along $\hat{j}$ with $j$ switching between $x$ and $y$ is consistent with the approach outlined in Ref.~\citenum{miao2019ExtensionFewestSwitches}.}
\label{fig:FIGURE_1}
\end{figure}

In the following, we assess the accuracy of this rescaling direction within FSSH. A comprehensive review of the application of the FSSH algorithm to topologically-trivial systems can be found elsewhere \cite{tully1990MolecularDynamicsElectronic, hammes-schiffer1994ProtonTransferSolution}. For topologically-nontrivial systems, the contribution of the diagonal elements of the derivative coupling to the classical momentum gives rise to a ``pseudomagnetic'' gauge field which yields an additional force in the classical equations of motion. A detailed derivation of these equations of motion can be found in the SM.

The avoided crossing governed by Eq.~\ref{eq:params_1} was previously investigated in detail in Ref.~\citenum{miao2019ExtensionFewestSwitches} where FSSH was applied while determining rescaling directions based on whichever gauge maximized the real part of the derivative coupling at that point in the trajectory. While this strategy yielded promising results for some values of $B$ and $W$, it induces abrupt changes in the rescaling direction (between $\hat{x}$ and $\hat{y}$) when the magnitude of the geometric rotation is increased which, as we show below, introduces inaccuracies. In what follows, we will compare results from FSSH under $\hat{x}$ rescaling predicted by our theory against the variable rescaling direction from Ref.~\citenum{miao2019ExtensionFewestSwitches}, henceforth referred to as $\hat{j}$, and against exact quantum results obtained with the Fourier transform method \cite{kosloff1983FourierMethodSolution}.

Shown in Figure~\ref{fig:FIGURE_1} are results for $A=20$, $B=1.0$, and either $W=2.0$ or $W=3.0$. Here, the nuclear wave packet is initialized on the upper ($n=1$) diabatic surface as
 \begin{equation}
 \Psi(\vc{q})= e^{i\vc{q}\cdot\vc{p}_{\text{init}}}e^{-\vert \vc{q}-\vc{q}_{\text{init}}\vert^{2}},
 \end{equation}
while being centered at $\vc{q}_{\text{init}}=(-3,0)$ and while moving towards the avoided crossing with $\vc{p}_{\text{init}}=(p_{\text{init}}^{x},0)$. Here, $p_{\text{init}}^{x}$ is varied between $8$ to $24$ \footnote{FSSH trajectories are initialized by sampling both adiabatic surfaces so as to stochastically reproduce the initial diabatic populations.}. Note that the diabatic surfaces cross at $y=0$ while the adiabatic surfaces maintain a constant energy of $\pm A$ and so a fully adiabatic trajectory (with no hops) yields transmitted populations fully on the $n=0$ diabatic surface.

For $\hat{x}$ rescaling the transmitted populations, $\rho_{nn}$, show quantitative agreement with exact results for both $W=2.0$ and $W=3.0$ across the full range of initial momenta. Similarly, high accuracy is reached for the $x$ component of the transmitted momenta for both diabatic surfaces, underscoring the validity of exclusively rescaling in the $x$ direction. Trajectories that remain on the same adiabatic surface arrive on the $n=0$ diabatic surface with $p^{x}_{0}\approx p^{x}_\mathrm{init}$ as expected. The trajectories that arrive on the $n=1$ diabatic surface, however, acquire additional momentum due to the rescaling in the $x$ direction that occurs following a hop. 

The $y$ component of the transmitted momentum, $p^{y}_{n}$, exhibits a moderate deviation from the exact result that can be attributed to a breakdown of the FSSH algorithm itself in the presence of pseudomagnetic forces, rather than the choice of $\hat{x}$ rescaling. Indeed, the underestimation of $p^{y}_{0}$ is due to trajectories hopping twice in the crossing region (therefore arriving on $n=0$) and experiencing the opposite pseudomagnetic force on the other adiabatic surface, reducing the net transmitted momentum from the expected result of $p^{y}_{0}=W$ (which was derived analytically in Ref.~\citenum{miao2019ExtensionFewestSwitches}).

Similarly, $p^{y}_{1}$ is underestimated due to the randomized location of the hops in the crossing region. In the SM we demonstrate that these sources of inaccuracy can be controlled by confining hops to either the $x>0$ or the $x<0$ region, finding an overestimation of $p^{y}_{1}$ for $x>0$ and an underestimation of $p^{y}_{1}$ for $x<0$. (In both cases $p^{y}_{0}$ yields near quantitative agreement with exact results due to the reduced number of trajectories that hop twice, confirming that the observed inaccuracies are again due to a breakdown of FSSH in the presence of pseudomagnetic fields instead of the choice of $\hat{x}$ rescaling.) Notably, if the hops were constrained to the $x=0$ point the trajectories would arrive with exactly $p^{y}_{1}=0$, having experienced the opposite pseudomagnetic forces on both surfaces and thereby acquiring no net momentum in the $y$ direction. As such, the surface hopping technique proposed by Tully and Preston \cite{tully1971TrajectorySurfaceHopping}, where hops only occur at avoided crossings, would not suffer from this source of inaccuracy \cite{miao2019ExtensionFewestSwitches}.

$\hat{j}$ rescaling, on the other hand, introduces further qualitative differences from exact results. For the applied parameters, $\hat{j}=\hat{y}$ when $\vert x\vert\leq 0.48$ ($\vert x\vert \leq 1.21$) for $W=2.0$ ($W=3.0$), and $\hat{j}=\hat{x}$ otherwise. Notably, this results in an overestimation of the population transfer between adiabatic surfaces, especially for $W=3.0$. When only considering the populations at $W=2.0$, it may seem that both $\hat{j}$ and $\hat{x}$ rescaling produce very similar results, but these results mask that the wavepacket evolves quite differently due to differences in momenta. For example, $\hat{j}$ rescaling yields an underestimation of $p^{x}_{1}$ as a consequence of instances of rescaling in the $y$ direction. For the same reason $p^{y}_{1}$ is significantly overestimated. Interestingly, $p^{y}_{0}$ incidentally agrees with exact results because rescaling in the $y$ direction inhibits trajectories from hopping twice due to insufficient momentum. Overall, however, the deviations introduced by $\hat{j}$ rescaling lead to qualitatively incorrect dynamics of the wavepacket.

While the rescaling procedure outlined in Ref.~\citenum{miao2019ExtensionFewestSwitches} is found to disagree with exact results for the parameters used here, it was previously shown to yield good agreement with exact results for a wide range of parameter choices \cite{miao2019ExtensionFewestSwitches}. Notably, for all of these cases this procedure predicts a rescaling along $\hat{x}$, and as such, the approach introduced in the present Letter (invariably predicting $\hat{x}$ rescaling) serves as an equally-viable means of momentum rescaling in these instances, but one that is more broadly generalizable.

\emph{Outlook.}--- In addition to finding the correct rescaling direction, our approach is fully consistent with our previous study \cite{krotz2022ReciprocalspaceFormulationSurface} which found that truncated reciprocal-space Hamiltonians may yield complex-valued derivative couplings. In this case, a truncated discrete Fourier transform can be applied to bring the system into a preferred basis in order to find real-valued derivative couplings. This is consistent with the lack of geometric phase effects in the equivalent real-space Hamiltonian \cite{krotz2021ReciprocalspaceFormulationMixed}. We also note possibilities for applying our approach to \emph{ab initio} trajectories of arbitrary dimension through the use of a diabatization scheme, relying on a vast body of previous works \cite{subotnik2008ConstructingDiabaticStates,subotnik2015RequisiteElectronicStructure,baer1976AdiabaticDiabaticRepresentations,mead1982ConditionsDefinitionStrictly,vanvoorhis2010DiabaticPictureElectron,guo2016AccurateNonadiabaticDynamics,werner1981MCSCFStudyAvoided,hoyer2016DQDQFElectronic}, and where the survey of the metric across diabatic bases will be instrumental.


\emph{Acknowledgements.}---This work was supported by the National Science Foundation under Grant No.~CHE-2145433.

\bibliography{Bibliography}
\end{document}


\title{Supplemental Material: Treating geometric phase effects in nonadiabatic dynamics}
\author{Alex Krotz}
\author{Roel Tempelaar}
\email{roel.tempelaar@northwestern.edu}
\maketitle

\tableofcontents

\section{Construction of a generic eigenvector matrix from the generators of SU($N$)}

In its most general form an eigenvector matrix $V_{0}$ is an element of U($N$) by virtue of its unitarity. By the multiplicativity of the determinant, $V_{0}$ can be made into an element of SU($N$) through a gauge transformation that leaves $H_{0}$ unchanged. It therefore suffices to construct a generic eigenvector matrix $V$ as an element of SU($N$), i.e.,
 \begin{align}
 V(\vc{I},\vc{R},\vc{G}) = \prod_{n}e^{I_{n}T^{I}_{n}}\prod_{n}e^{R_{n}T^{R}_{n}}\prod_{n}e^{G_{n}T^{E}_{n}},\label{eq:V_1}
 \end{align}
 where $\vc{I}=(I_{1},I_{2},...,I_{N'})$, $\vc{R}=(R_{1},R_{2},...,R_{N'})$, and $\vc{G}=(G_{1},G_{2},...,G_{N-1})$ are functions associated with the subsets $\{T^{I}_n\}$, $\{T^{R}_n\}$, and $\{T^{E}_n\}$, respectively.
 
The ordering of generators in Eq.~\ref{eq:V_1} is taken as a convention and the inclusion of the terms associated with $\{T^{E}_n\}$ in Eq.~\ref{eq:V_1} is an intentional one meant to emphasize their role as gauge transformations. Variations in $\vc{G}$ leave the Hamiltonian unchanged and so these terms can be discarded. Therefore, instead of SU($N$) we represent $V$ by a subset of SU($N$) which still accounts for every possible Hamiltonian,
 \begin{align}
 V(\vc{I},\vc{R})&=\prod_{n}e^{I_{n}T^{I}_{n}}\prod_{n}e^{R_{n}T^{R}_{n}}.
 \end{align}
 
\section{Construction of $H$ from the generators of SU($2$)}

For $N=2$, $H$ can be constructed by employing the generators of SU($2$) given by
 \begin{align}
 T^{E}=\left(\begin{array}{cc}
 i & 0 \\
 0 & -i
 \end{array}\right),
 \quad
 T^{R} = \left(\begin{array}{cc}
 0 & -1 \\
 1 & 0
 \end{array}\right),
 \quad
 T^{I} = \left(\begin{array}{cc}
 0 & i \\
 i & 0 
 \end{array}\right).
 \end{align}
 A generic eigenvector matrix can be written following Eq.~2,
 \begin{align}
 V(I,R)=e^{IT^{I}}e^{RT^{R}}.
 \end{align}
The eigenvalue matrix, on the other hand follows from Eq.~1,
 \begin{align}
 E(\lambda)=\left(\begin{array}{cc}
 -\lambda & 0 \\
 0 & \lambda
 \end{array}\right),
 \end{align}
 upon which the Hamiltonian follows from Eq.~3 as
 \begin{align}
 H(\lambda,I,R)&=V(I,R)E(\lambda)V^{\dagger}(I,R)\nonumber\\
 &=\lambda \left(\begin{array}{cc}
 -\cos(2 R)\cos(2 I) & -\sin(2 R)+i\cos(2R)\sin(2I) \\
 -\sin(2 R)-i\cos(2R)\sin(2I) & \cos(2 R)\cos(2 I)
 \end{array}\right).
 \end{align}

\section{Survey of diabatic bases}
\begin{figure}
 \centering
 \includegraphics{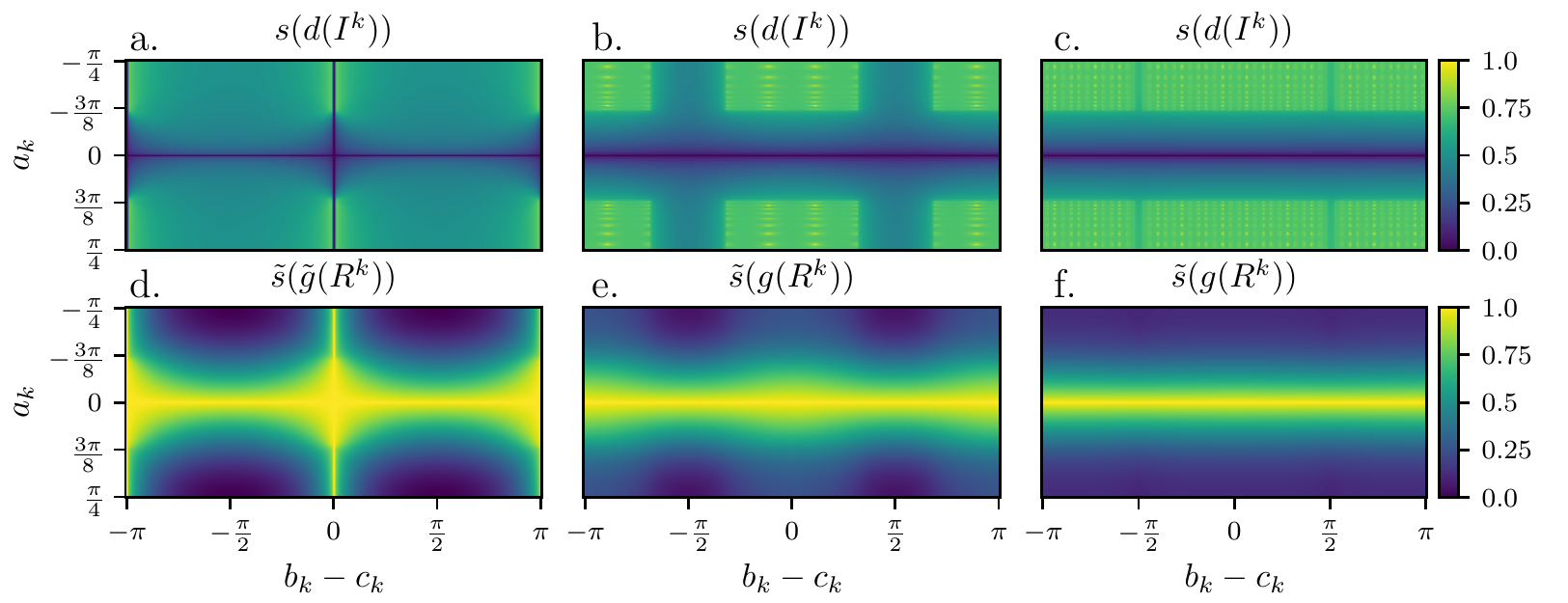}
 \caption{The metric to second order and aggregate rescaling with respect to $\hat{x}$ for $B=1.0$ with $W=0.0$ (a,d), $W=2.0$ (b,e), and $W=3.0$ (c,f). Panels a, b, and c all exhibit minima at points where panels d, e, and f exhibit maxima equal to $1$, respectively, meaning that the rescaling direction for the preferred diabatic bases all correspond to $\hat{x}$.}
 \label{fig:SM_1}
 \end{figure}

Here we evaluate the metric $d(I^{k})$ for the avoided crossing considered in the main text by scanning over all possible diabatic bases. We recognize that because any $V(R^{k},I^{k})$ is an element of SU($N$), all relevant diabatic basis transformations are also elements of SU($N$) and can thus be assumed to take the form
 \begin{align}
 D_{k} = \left(\begin{array}{cc}
 e^{i b_{k}}\cos a_{k} & -e^{-i c_{k}}\sin a_{k} \\
 e^{i c_{k}}\sin a_{k} & e^{-i b_{k}}\cos a_{k} 
 \end{array}\right)e^{-\frac{\pi}{4}T^{R}},\label{eq:D_k}
 \end{align} 
where $a_{k}\in[0,\pi)$ and $b_{k},c_{k}\in[0,2\pi)$. The constant term $e^{-\frac{\pi}{4}T^{R}}$ conveniently enables the metric to be written in terms of $a_{k}$ and $b_{k}-c_{k}$, facilitating the visualization of the results in Fig.~\ref{fig:SM_1}. 

In terms of the Hamiltonian in the diabatic basis given in the main text (Eq.~8) we can write $I^{k}$ as
 \begin{align}
 I^{k}=\frac{1}{2}\arctan\left(\frac{\Im([H^{k}_{0}]_{12})}{\Re([H^{k}_{0}]_{22})}\right),
 \end{align}
where $[H^{k}_{0}]_{12}$ is the off-diagonal matrix element of $H^{k}_{0}$ and $[H^{k}_{0}]_{22}$ is the lower diagonal element. The metric to second order follows as
 \begin{align}
 d(I^{k})=\Omega^{-1}\int \left[\left(\frac{\partial I^{k}}{\partial x}\right)^{2} + \left(\frac{\partial I^{k}}{\partial y}\right)^{2} + \frac{1}{4}\left(\frac{\partial^{2}I^{k}}{\partial x^{2}}\right)^{2}
 + \frac{1}{4}\left(\frac{\partial^{2}I^{k}}{\partial y^{2}}\right)^{2} + \frac{1}{2}\left(\frac{\partial^{2}I^{k}}{\partial x\partial y} \right)^{2}\right]^{\frac{1}{2}}d\,q.
 \end{align} 

For $N=2$ the purely-real contribution to Eq.~4 is given by $\tilde{V}^{\dagger}(R^{k})\nabla\tilde{V}(R^{k})=\nabla R^{k} T^{R}$, and so the momentum rescaling direction is the $\vc{q}$-dependent vector $\nabla R^{k}$. To quantify deviations of $\nabla R^{k}$ from the $x$ direction at the aggregate level, we construct the quantity 
 \begin{align}
 g(R^{k})=\Omega^{-1}\int\left(\hat{x}\cdot\frac{ \nabla R^{k}}{\vert\nabla R^{k}\vert}\right)^{2}d\,q
 \end{align}
where $\hat{x}$ is the unit vector in the $x$ direction and $\Omega$ is the area integrated. $g(R^{k})=1$ when $\nabla R^{k}$ is aligned with $\hat{x}$ and $g(R^{k})<1$ when the direction of $\nabla R^{k}$ deviates from $\hat{x}$. To compare the rescaling directions of the $W=0.0$ (topologically trivial) case, we instead make use of the quantity
 \begin{align}
 \tilde{g}(R^{k})=(\gamma\Omega)^{-1}\int\left(\hat{x}\cdot \nabla R^{k}\right)^{2}d\,q,
 \end{align}
where $\gamma$ is a normalization constant such that $\max_{k}(\tilde{g}(R^{k}))=1$ (where $\max_k$ is the maximum with respect to the diabatic basis index $k$). This choice of normalization is for ease of visualization for the topologically trivial case where the singular gauge yields $\tilde{g}(R^{k})=0$ while being undefined for $g(R^{k})$. 

In Fig.~\ref{fig:SM_1} we compare the metric to second order and aggregate rescaling direction with respect to $\hat{x}$ for $B=1.0$ and $W=0.0$, $W=2.0$, and $W=3.0$. We make use of the scaling functions 
 \begin{align}
 s(d(I^{k}))=\frac{1}{\sqrt{2}}\left[\log_{10}\left(d(I^{k}) - \mathrm{min}_{k}(d(I^{k}))+1\right)\right]^{\frac{1}{4}}
 \end{align}
 and 
 \begin{align}
 \tilde{s}(g(R^{k}))=1-\frac{2}{\pi}\cos^{-1}(g(R^{k}))
 \end{align}
for ease of visualization of the minima and maxima of the metric and aggregate rescaling, respectively. Notably, $\tilde{s}(g(R^{k}))=1$ indicates that $\nabla R^{k}$ and $\hat{x}$ are in the same direction, and $s(d(I^{k}))=0$ implies that $d(I^{k})$ is minimized.
 
As stated in the main text, the minimum metric is found to be $d(I_{\bar{k}})=\frac{1}{2}\vert W\vert$, which is invariant to the area integrated and which truncates after first order. Therefore, in order confirm that this metric is indeed a minimum by checking that all other metrics are greater than (or equal to) $d(I^{\bar{k}})$, these other metrics need only be evaluated up to second order. Integrals are performed over $x\in [-0.5,0.5]$ and $y\in[-0.5,0.5]$, each sampled with a resolution of $100$ points. Much like the truncation of the metric to second order the region of integration need only be sufficiently large to establish a lower bound for the metric. 

The scan over diabatic bases is performed by discretizing $a_{k}\in[-\frac{\pi}{4},\frac{\pi}{4}]$ and $b_{k}\in[-\pi,\pi]$ with a resolution of $161$ points each and with $c_{k}=0$ which is a subset of the possible diabatic bases. Because $\nabla I^{k}$ and $R^{k}$ are $\pi$-periodic in $a_{k}$ (changing sign with period $\frac{\pi}{2}$ in $a_{k}$) and $2\pi$-periodic in $b_{k}-c_{k}$ and higher order terms of the metric retain this periodicity, the metric and aggregate rescaling direction are both $\frac{\pi}{2}$-periodic in $a_{k}$ and $2\pi$-periodic in $b_{k}-c_{k}$. Thus, evaluating $d(I^{k})$ over this subset of diabatic bases is sufficient to find every unique combination of the metric and aggregate rescaling. By recognizing that $d(I^{k})$ is minimized in the same bases where $\tilde{s}(g(R^{k}))=1$, it is clear that all preferred bases yield $\hat{x}$ rescaling. 
 
\section{Derivation of classical equations of motion}

Trajectory surface hopping techniques assume adiabatic evolution of a quantum state on a single adiabatic surface in between hops. This corresponds to the Born-Huang approximation under which the classical Hamiltonian takes the form \cite{mead1992GeometricPhaseMolecular}
 \begin{align}
 H^{\mathrm{cl}}=\frac{1}{2}(\tilde{\vc{p}}-\vc{A}_{\alpha})^{2} + V_{\alpha}(\vc{q}).
 \end{align}
 Here, $\tilde{\vc{p}}$ is the gauge-dependent canonical momentum and $\vc{A}_{\alpha}\equiv i\langle\alpha\vert\nabla\alpha\rangle$ is the gauge potential. Classical mechanics involves the kinetic momentum $\vc{p}=\tilde{\vc{p}}-\vc{A}_{\alpha}=\dot{\vc{q}}$, where the second equality assumes all classical masses to be set to unity. Note that in the main text it is the kinetic momenta that are sampled to represent the nuclear wave packet and the canonical momenta are never explicitly utilized. This approach is consistent with how FSSH is ordinarily applied, seeing that for topologically trivial systems the difference between the kinetic and canonical momenta vanishes in the appropriate gauge.

The equations of motion then follow as
 \begin{align}
 \dot{q}_{i}=\frac{\partial H^\mathrm{cl}}{\partial p_{i}}
 \end{align}
 and
 \begin{align}
 \dot{p}_{i} &= -\frac{\partial H^{\mathrm{cl}}}{\partial q_{i}}-\frac{d A_{i}}{dt}\nonumber\\
 &= \sum_{k}\left(\tilde{p}_{k}-A_{\alpha}^{k}\right)\frac{\partial A_{\alpha}^{k}}{\partial q_{i}}-\frac{\partial A_{\alpha}^{i}}{\partial q_{k}}\dot{q}_{k} - \frac{\partial V_{\alpha}}{\partial q_{i}}\nonumber\\
 &=\sum_{k}p_{k}\left(\frac{\partial A^{k}_{\alpha}}{\partial q_{i}}-\frac{\partial A^{i}_{\alpha}}{\partial q_{k}}\right)-\frac{\partial V_{\alpha}}{\partial q_{i}},\label{eq:pdot_1}
 \end{align}
where we made use of the chain rule to obtain the gauge-invariant pseudomagnetic force terms, which are reminiscent of the Lorentz force on a charged particle in a magnetic field.

\section{Influence of stochastic hopping on $p^{y}_{n}$}

 \begin{figure}
 \centering
 \includegraphics{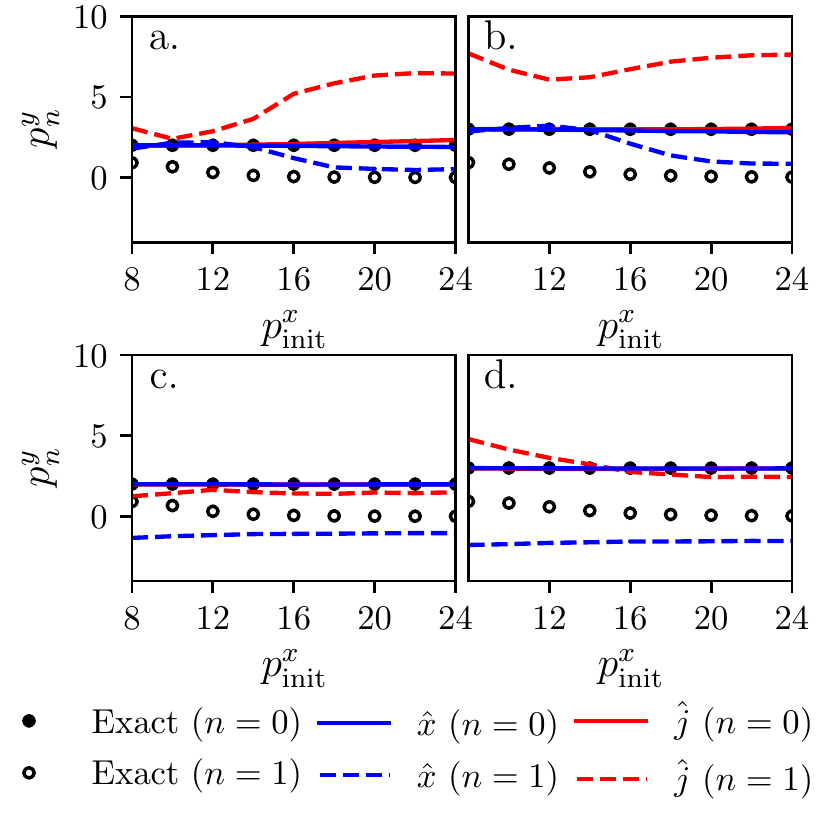}
 \caption{Transmitted momentum in the $y$ direction for $B=1.0$ and $W=2.0$ (a,c) or $W=3.0$ (b,d) for FSSH where hops are allowed only if $x>0$ (a-b) or $x < 0$ (c-d). }
 \label{fig:SM_2}
 \end{figure}
 
Fig.~\ref{fig:SM_2} shows the $y$-component of the transmitted momenta where hops have been restricted to occurring in either the $x>0$ or the $x < 0$ regions. For the latter, $p^{y}_{1}$ is more positive due to trajectories experiencing more of the pseudomagnetic force on the upper adiabatic surface. For hops restricted to $x > 0$, however, trajectories experience the opposite pseudomagnetic force on the lower adiabatic surface leading to a negative $p^{y}_{1}$. As mentioned in the main text, the improved agreement with exact results for $p^{y}_{0}$ upon restricting the region in which hops may occur is reflective of the reduction in trajectories that hop twice. In the case of $\hat{x}$ rescaling this is a direct consequence of restricting the hops while for $\hat{j}$ rescaling it is due to insufficient momentum available in the $y$ direction for a hop to occur. 

\bibliography{Bibliography}